\documentclass[prl,twocolumn,preprintnumbers,nofootinbib,amsmath,amssymb]{revtex4-1}
\usepackage{graphicx,epsfig,slashed,booktabs,bm,bbm,psfrag}

\def\Tr{{\rm Tr}}
\def\lQ{\Lambda_{\rm QCD}}
\def\als{\alpha_s}

\begin{document}

\title{Quarkonium suppression in heavy-ion collisions:\\ an open quantum system approach}
\author{Nora Brambilla}
\affiliation{Physik Department, Technische Universit\"at M\"unchen, D-85748 Garching, Germany}
\affiliation{Institute for Advanced Study, Technische Universit\"at M\"unchen, Lichtenbergstrasse 2 a, D-85748 Garching, Germany}
\author{Miguel A.~Escobedo}
\affiliation{Department of Physics, P.O. Box 35, FI-40014 University of Jyv\"askyl\"a, Finland} 
\affiliation{Institut de Physique Th\'eorique, Universit\'e Paris Saclay, CNRS, CEA, F-91191 Gif-sur-Yvette, France}
\author{Joan Soto}
\affiliation{Departament de F\'\i sica Qu\`antica i Astrof\'\i sica and Institut de Ci\`encies del Cosmos, 
Universitat de Barcelona, Mart\'\i $\;$ i Franqu\`es 1, 08028 Barcelona, Catalonia, Spain}
\author{Antonio Vairo}
\affiliation{Physik Department, Technische Universit\"at M\"unchen, D-85748 Garching, Germany}

\date{\today}

\preprint{ICCUB-16-044, TUM-EFT 55/14}

\begin{abstract}
We address the evolution of heavy-quarkonium states in an expanding quark-gluon plasma 
by implementing effective field theory techniques in the framework of open quantum systems.
In this setting we compute the nuclear modification factors for quarkonia 
that are $S$-wave Coulombic bound states in a strongly-coupled quark-gluon plasma.
The calculation is performed at an accuracy that is leading-order in the heavy-quark density expansion and next-to-leading order in the multipole expansion. 
The quarkonium density-matrix evolution equations can be written in the Lindblad form, and, hence, they account for both dissociation and recombination.
Thermal mass shifts, thermal widths and the Lindblad equation itself depend on only two non-perturbative parameters: the heavy-quark momentum diffusion coefficient 
and its dispersive counterpart. Finally, by numerically solving the Lindblad equation, 
we provide results for the $\Upsilon(1S)$ and $\Upsilon(2S)$ suppression.
\end{abstract}

\maketitle

\section{Introduction}
Quarkonium suppression has been since long suggested as a hard probe for the quark-gluon plasma formed in high-energy heavy-ion collisions~\cite{Matsui:1986dk}. 
Experimentally, quarkonium provides a potentially clean signal through dilepton decays~\cite{McLerran:1984ay}. 
Theoretically, it allows the use of non-relativistic effective field theories to factorize high-energy contributions from low-energy ones. 
The latter can be eventually computed by lattice QCD~\cite{Brambilla:2004wf,Brambilla:2010cs,Brambilla:2014jmp}.
Nevertheless, many processes contribute to the final heavy-quarkonium observables, 
among these the hydrodynamic evolution of the plasma, and quarkonium production, dissociation and regeneration in the different medium conditions.

The problem may be simplified if one considers the quarkonium ground state and particularly the bottomonium $1S$ states.
(To a lesser extent the same may apply to the charmonium ground state and to the bottomonium $2S$ states.)
In this case not only one can argue that the mass of the heavy quark, $m$, is the largest scale of the problem, which qualifies the system as non-relativistic, 
but also that the typical momentum transfer between the heavy quarks is the next relevant scale for a certain range of temperatures, which qualifies the system as Coulombic.
The Bohr radius of such a system, $a_0 \sim 1/(m \als)$ (the inverse of the typical momentum transfer), is the scale at which the strong coupling is computed in the potential. 
The Coulomb potential of a quark-antiquark pair in a color-singlet configuration reads, $V_s(r) = - C_F\als/r$,
and in a color-octet configuration, $V_o(r) = \als/(2N_cr)$; $N_c=3$ is the number of colors and $C_F = (N_c^2-1)/(2N_c) = 4/3$ is
the Casimir of the fundamental representation of SU(3).

At least two mechanisms of quarkonium decay in a medium have been identified over the years:
gluodissociation~\cite{Kharzeev:1994pz,Xu:1995eb} and dissociation via parton scattering~\cite{Grandchamp:2001pf,Grandchamp:2002wp}.
Both dissociation mechanisms have been studied in an effective field theory framework applied to Coulombic heavy quark-antiquark states in a weakly-coupled plasma 
in~\cite{Brambilla:2008cx,Brambilla:2010vq,Vairo:2010bm,Brambilla:2011sg,Brambilla:2013dpa,Escobedo:2013tca}.
An extensive phenomenological analysis of bottomonium suppression can be found in~\cite{Krouppa:2016jcl,Krouppa:2015yoa,Strickland:2011aa}.

In a weakly-coupled plasma one assumes the hierarchy $\pi T \gg m_D$,  
where $T$ is the temperature and $m_D \sim  gT$ the Debye mass of the plasma
(the factor $\pi$ is a remnant of the Matsubara frequencies).
The theoretical advantage from this situation is that one may use perturbation theory as a computational tool. 
It is uncertain, however, if a weakly-coupled plasma is what best describes the medium formed in heavy-ion collisions at the LHC.
A more conservative approach consists in assuming that the plasma is strongly coupled, i.e., $\pi T \sim m_D$. 
This is the situation that we will analyze in the following, 
where we will assume the hierarchy of scales  
\begin{equation}
m \gg \frac{1}{a_0} \sim m\als \gg \pi T \sim m_D \gg \hbox{any other scale}\,.
\label{hierarchy}
\end{equation}
The other scales include the binding energy and $\lQ$, whose relative size is not specified.
We will compute the quarkonium thermal decay width and mass shift,  
write and solve the heavy quark-antiquark evolution equations, 
and finally evaluate the bottomonium nuclear modification factor.

Bottomonium suppression has been measured by CMS at 2.76~TeV~\cite{Chatrchyan:2012lxa}. Analyses for LHC data at 5.02~TeV are under way
(see, e.g.,~\cite{Fronze:2016}). Recently also the STAR experiment has considered bottomonium suppression data at the much lower energy 
of 193~GeV improving on previous studies by the PHENIX collaboration~\cite{Adamczyk:2016dzv}.

\section{Quarkonium decay width and mass shift in a strongly-coupled plasma}
Under the condition \eqref{hierarchy}, the effective field theory suited to describe heavy quark-antiquark pairs 
at an energy scale lower than $m\als$ but larger than the thermal scales is potential non-relativistic QCD (pNRQCD)~\cite{Pineda:1997bj,Brambilla:1999xf,Brambilla:2004jw}.
According to the hierarchy of energy scales, pNRQCD may be computed setting to zero the temperature and any other scale lower than $m\als$~\cite{Brambilla:2008cx}.
The remaining scales provide non-perturbative contributions: contributions from thermal scales 
have to be resummed to all orders because they are induced by a strongly-coupled plasma, but also contributions scaling with the 
binding energy of the system may be non-perturbative if the binding energy is not larger than~$\lQ$.

The Lagrangian of pNRQCD at next-to-leading order in the multipole expansion reads 
\begin{eqnarray}
&& \!\! \mathcal{L}_{\rm pNRQCD}  =  \int d^3r \; 
\Tr\left[{\rm S}^\dagger\left(i\partial_0-h_s\right) {\rm S} + {\rm O}^\dagger\left(iD_0-h_o\right) {\rm O}\right]
\nonumber \\
&& 
\quad 
+ \Tr\left[  \left({\rm O}^\dagger{\bf r}\cdot g{\bf E} {\rm S} + \hbox{H.c.}\right)
+\frac{1}{2} \left({\rm O}^\dagger{\bf r}\cdot g{\bf E} {\rm O} + \hbox{c.c.}\right)  \right]
\nonumber \\
&& \quad 
+\mathcal{L}_{\rm light} \,,
\label{eq:Lagrangian_pNRQCD}
\end{eqnarray}
where $r$ is the distance between the heavy quark and the antiquark (the above Lagrangian is accurate up to order~$r$),
${\rm S} = S\,\mathbbm{1}_c/\sqrt{N_c}$ and ${\rm O}=\sqrt{2}O^aT^a$ stand for the heavy quark-antiquark fields in a color-singlet and a color-octet configuration respectively,
$h_s = {\bf p}^2/m + V_s$ is the color-singlet Hamiltonian, $h_o = {\bf p}^2/m + V_o$ is the color-octet Hamiltonian (${\bf p} = -i {\bm\nabla}_r$), 
${\bf E}$ is the chromoelectric field, $g$ the strong coupling and H.c. and c.c. stand for Hermitian conjugate and charge conjugate respectively. 
The term $\mathcal{L}_{\rm light}$ denotes the QCD Lagrangian with light quarks only.
The covariant derivative acting on the octet field ${\rm O}$ in \eqref{eq:Lagrangian_pNRQCD} 
can be eliminated by means of suitable field redefinitions: ${\rm O} \to \Omega {\rm O} \Omega^\dagger$ and ${\bf E} \to \Omega {\bf E} \Omega^\dagger$,
where $\displaystyle \Omega = {\rm P} \, \exp\left[ -ig \int_{-\infty}^t ds \, A_0(s,{\bf R}) \right]$ and ${\bf R}$ is the center of mass coordinate.
We use these field redefinitions when deriving the evolution equations in the next section.

At order $r$ in the multipole expansion quark-antiquark pairs interact with the medium through chromoelectric-dipole interactions.
In \eqref{eq:Lagrangian_pNRQCD} we have neglected radiative corrections to the Wilson coefficients of these terms, since they are beyond our accuracy.
According to \eqref{hierarchy} the thermodynamical scales are much larger than the binding energy.
Hence, the latter can be neglected in the color-singlet self energy, which reads 
\begin{eqnarray}
\Sigma_s(t) &=& \frac{g^2}{6N_c}\,r^2 \int_{t_0}^{t} dt_2 \, \langle E^{a,i}(t,{\bf 0})E^{a,i}(t_2,{\bf 0})\rangle \,,
\label{eq:sigma}
\end{eqnarray}
where $\langle \cdots \rangle$ stands for the thermal average.
The time $t_0$ is the formation time of the quark-gluon plasma, i.e., the initial time for the evolution of the heavy quark-antiquark pairs in the medium. 

In the following, we will assume that $t-t_0$ is larger than any other time scale of the system 
and that the evolution of the temperature is quasistatic: $1/T \times dT/dt \sim 1/t \lesssim \hbox{binding energy}$.
As a consequence of these two assumptions 
we can approximate $\displaystyle \int_{t_0}^tdt_2  \,  \langle E^{a,i}(t,{\bf 0})E^{a,i}(t_2,{\bf 0})\rangle 
\approx \int_{-\infty}^{+\infty} \!\!\! ds \,  \langle \,{\rm T}\, E^{a,i}(s,{\bf 0})E^{a,i}(0,{\bf 0})\rangle/2$. 
The last correlator is the time-ordered one. 
It may be computed as if the medium was in thermal equilibrium at a slowly varying temperature $T$.

The leading contribution to the thermal decay width is given by the imaginary part of the color-singlet self energy.
Under the above assumptions, it reads for $1S$ states 
\begin{equation}
\Gamma =  -2 \langle {\rm Im}\,(-i\Sigma_s) \rangle = 3 a_0^2 \,\kappa \,.
\label{width}
\end{equation}
The heavy-quark momentum diffusion coefficient, $\kappa$, is defined as~\cite{CasalderreySolana:2006rq,CaronHuot:2007gq}
\begin{equation}
\kappa = \frac{g^2}{6\,N_c} \,  {\rm Re} \int_{-\infty}^{+\infty}ds \, \langle  \,{\rm T}\,E^{a,i}(s,{\bf 0}) E^{a,i}(0,{\bf 0})\rangle \,.
\label{kappa}
\end{equation} 
A recent lattice determination of $\kappa$ found~\cite{Francis:2015daa}: 
\begin{equation}
1.8 \lesssim \frac{\kappa}{T^3} \lesssim 3.4 \,.
\label{kapparange}
\end{equation}
This estimate has been obtained from a pure SU(3) plasma at a temperature of about 1.5~$T_c$.
With this value of $\kappa$, $m_b = 4.8$~GeV and $1/a_0 = 1.334$~GeV that follows from the self-consistency equation $1/a_0 = m_b C_F \als(1/a_0)/2$, 
we obtain for the $\Upsilon(1S)$: $3.0\,\hbox{GeV}^{-2} \, T^3 \lesssim \Gamma_{\Upsilon(1S)} \lesssim 5.7 \,\hbox{GeV}^{-2} \, T^3$.
This provides a thermal width around 100~MeV for temperatures of about 300~MeV.
We recall that the cross-over temperature to the quark-gluon plasma, $T_c$, as measured by lattice QCD is about 150~MeV~\cite{Borsanyi:2010bp,Bazavov:2014pvz,Bazavov:2016uvm}.

The leading contribution to the quarkonium mass shift is given by the real part of the color-singlet self energy.
For $1S$ states it reads 
\begin{equation}
\delta m =  \langle {\rm Re}\,(-i\Sigma_s) \rangle = \frac{3}{2} a_0^2 \,\gamma \,, 
\label{mass}
\end{equation}
where 
\begin{equation}
\gamma = \frac{g^2}{6\,N_c} \, {\rm Im}\int_{-\infty}^{+\infty}ds \, \langle \,{\rm T} \,E^{a,i}(s,{\bf 0}) E^{a,i}(0,{\bf 0})\rangle
\,.
\label{gamma}
\end{equation} 
So far the only estimate that we have for $\gamma$ is the perturbative calculation done at leading order in~\cite{Brambilla:2008cx}:
\begin{equation}
\gamma = -  2 \zeta(3) \, C_F \left(\frac{4}{3}N_c+n_f\right) \als^2\, T^3 \, ,
\end{equation}
where $n_f$ is the number of light quarks.
In the case of a strongly-coupled plasma, however, there is no reason to believe that perturbation theory provides a good approximation of $\gamma$.
For example, the leading-order estimate of $\kappa$ is far off the non-perturbative result~\cite{Banerjee:2011ra}.

\section{Evolution equations}
The yield of quarkonium $nS$ states in heavy-ion collisions normalized with respect to the yield in $pp$ collisions, as measured 
from dilepton decays, is called the quarkonium nuclear modification factor,~$R_{AA}(nS)$. 
It can be expressed as the density of (color singlet) $nS$ states in heavy-ion collisions 
normalized with respect to the same quantity in $pp$ collisions~\cite{McLerran:1984ay,Brambilla:2016fire}.
The density of color-singlet heavy quark-antiquark states, $\rho_s$, and color-octet ones, 
$\rho_o^{ba} =  \rho_o \, \delta^{ab}/(N_c^2-1)$,  
may be expressed in the close-time-path formalism as a singlet and octet propagator respectively 
that propagate from the upper branch (labeled 1) to the lower branch (labeled 2) of the time path~\cite{Berges:2004yj}:
$\langle {\bf r}'|\rho_s(t;t)|{\bf r}\rangle$  $=  \Tr\{ \rho \, S^\dagger(t, {\bf r}, {\bf R}) S(t, {\bf r}', {\bf R})\}$ 
$=  \langle S_1(t,{\bf r}',{\bf R}) S_2^\dagger(t,{\bf r},{\bf R})\rangle$,  
$\langle {\bf r}'|\rho_o(t;t)|{\bf r}\rangle\delta^{ab}/(N_c^2-1)$ 
$= \Tr\{ \rho \, O^{a\dagger}(t, {\bf r}, {\bf R}) O^b(t, {\bf r}', {\bf R})\}$ 
$=  \langle O_1^b(t,{\bf r}',{\bf R}) O_2^{a\dagger}(t,{\bf r},{\bf R})\rangle$.
We have assumed that the heavy quarks comove with the medium, so that we do not need to consider the center of mass motion.

The evolution equations for the singlet and octet densities may be computed, therefore, by considering 
self-energy contributions similar to the one considered in the previous section, 
but now describing the evolution of the $\langle S_1 S_2^\dagger \rangle$ and $\langle O_1^b O_2^{a\dagger} \rangle$ propagators. 
Keeping only terms linear in the heavy-quark densities and resumming self-energy contributions 
by means of a Schwinger--Dyson equation, we obtain the evolution equations 
\begin{eqnarray}
\frac{d\rho_s(t;t)}{dt} &=& -i[h_s,\rho_s(t;t)] - \Sigma_s(t)\rho_s(t;t) - \rho_s(t;t)\Sigma_s^\dagger(t)
\nonumber\\
&& +\Xi_{so}(\rho_o(t;t),t)\,,
\label{eq:ev} \\
\frac{d\rho_o(t;t)}{dt} &=& -i[h_o,\rho_o(t;t)] - \Sigma_o(t)\rho_o(t;t) - \rho_o(t;t)\Sigma_o^\dagger(t)
\nonumber\\
&& +\Xi_{os}(\rho_s(t;t),t)+\Xi_{oo}(\rho_o(t;t),t)\,,
\label{eq:ev_octet}
\end{eqnarray}
where, assuming the hierarchy of energy scales \eqref{hierarchy} that allows to neglect energy-dependent exponentials, we have 
\begin{eqnarray}
\Sigma_s(t) &=& \frac{r^2}{2}\left[ \kappa(t) + i \gamma(t) \right]\,,
\label{strong1}
\\
\Sigma_o(t) &=& \frac{N_c^2-2}{2(N_c^2 -1)}\frac{r^2}{2}\left[ \kappa(t) + i \gamma(t) \right]\,,
\label{strong2}
\\
\Xi_{so}(\rho_o,t) &=& \frac{1}{N_c^2 -1}\,r^i \, \rho_o \, r^i\,\kappa(t)\,,
\label{strong3}
\\
\Xi_{os}(\rho_s,t) &=& r^i \, \rho_s \, r^i\,\kappa(t)\,,
\label{strong4}
\\
\Xi_{oo}(\rho_o,t) &=& \frac{N_c^2-4}{2(N_c^2 -1)}\,r^i \, \rho_o \, r^i\,\kappa(t)\,.
\label{strong5}
\end{eqnarray}

The equations show the coupled evolution of the singlet and octet densities.
Their interpretation is straightforward:
the function $\Xi_{so}$ accounts for the production (or regeneration) of singlets through the decay of octets, 
while the functions $\Xi_{os}$ and $\Xi_{oo}$ account for the production of octets through the decays of singlets and octets respectively. 
These two octet production mechanisms can be traced back to the two different sets of chromoelectric dipole operators in the pNRQCD Lagrangian \eqref{eq:Lagrangian_pNRQCD}. 

With the above functions, we can rewrite the evolution equations \eqref{eq:ev} and \eqref{eq:ev_octet} in the Lindblad form~\cite{Lindblad:1975ef,Gorini:1975nb}:
\begin{equation}
\frac{d\rho}{dt}=-i[H,\rho]+\sum_n \left( C_n\,\rho \,C_n^\dagger-\frac{1}{2}\{C_n^\dagger C_n,\rho\} \right)\,,
\label{eq:Lindblad}
\end{equation}
where $H$ is a Hermitian operator and the operators $C_n$ are called collapse operators.
In our case, the matrix $\rho$ is 
\begin{equation}
\rho=\left(\begin{array}{c c}
\rho_s & 0 \\
0 & \rho_o\end{array}\right)\,,
\end{equation}
the operator $H$ is 
\begin{equation}
H = \left(\begin{array}{c c}
h_s & 0\\
0 & h_o
\end{array}\right)
+ \frac{r^2}{2}\,\gamma(t)\, 
\left(\begin{array}{c c}
1 & 0\\
0 & \frac{N_c^2-2}{2(N_c^2-1)}
\end{array}\right)\,,
\end{equation}
and we need six collapse operators, 
\begin{equation}
C^0_i=\sqrt{\frac{\kappa(t)}{N_c^2-1}}\,r^i\left(\begin{array}{c c}
0 & 1\\
\sqrt{N_c^2-1} & 0
\end{array}\right)\,,
\end{equation}
\begin{equation}
C^1_i=\sqrt{\frac{(N_c^2-4)\kappa(t)}{2(N_c^2-1)}}\,r^i\left(\begin{array}{c c}
0 & 0\\
0 & 1
\end{array}\right)\,.
\end{equation}
The Lindblad equation has been studied in relation with quarkonium in a quark-gluon plasma also in~\cite{Akamatsu:2014qsa}.

\section{Bottomonium suppression}
We assume that the temperature of the quark-gluon plasma evolves slowly according to
\begin{equation}
T=T_0\left(\frac{t_0}{t}\right)^{v_s^2}\,,
\label{Tbj}
\end{equation}
where $T_0$ is the initial temperature and $v_s$ is the velocity of sound in the medium~\cite{Bjorken:1982qr}. 
In a deconfined plasma at a very high temperature $v_s^2= 1/3$. 
As values of $T_0$ and $t_0$ for central collisions at the LHC we take $T_0=475$~MeV and $t_0=0.6$~fm~\cite{Alberico:2013bza}.

\begin{table}[ht]
\begin{center}
\begin{tabular}{|c|c|c|}
\hline
centrality (\%) & $\langle b\rangle$ (fm) & $T_0$ (MeV) \\ 
\hline
$0-10$ & $3.4$ & $471$ \\
$10-20$ & $6.0$ & $461$ \\
$20-30$ & $7.8$ & $449$ \\
$30-50$ & $9.9$ & $425$ \\
$50-100$ & $13.6$ & $304$ \\
\hline
\end{tabular}
\end{center}
\caption{Initial temperature of the fireball, $T_0$, for different centrality bins 
(10\% centrality means that the 10\% most central collisions have been selected)
computed by means of the Glauber model from the mean impact parameters, $\langle b\rangle$, taken from~\cite{Chatrchyan:2011sx}. 
\label{tab:central}}
\end{table}

We study collisions with different centralities. 
For a plasma that is homogeneous and isotropic the only effect that a difference in centrality produces is a change in the initial value of the energy density and hence in $T_0$. 
The values of $T_0$ for different centralities and mean impact parameters that we use are listed in Tab.~\ref{tab:central}.

\begin{figure}[h]
\includegraphics[scale=0.35]{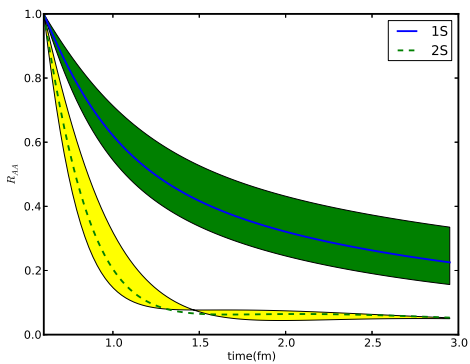}
\caption{Time evolution of $R_{AA}$ for $\kappa/T^3$ in the range \eqref{kapparange}, $\gamma = 0$ and $\delta =1$, 
for 30-50\% centrality.
\label{fig:ev1}
}
\end{figure}

\begin{figure}[h]
\includegraphics[scale=0.35]{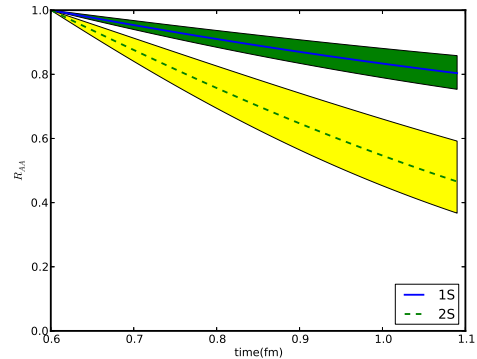}
\caption{Time evolution of $R_{AA}$ for $\kappa/T^3$ in the range \eqref{kapparange}, $\gamma = 0$ and $\delta =1$, 
for 50-100\% centrality.
\label{fig:ev2}
}
\end{figure}

According to \eqref{Tbj} and Tab.~\ref{tab:central}, if the evolution starts at $t_0=0.6$~fm,
the fireball cools down to about 250~MeV at about 4~fm for the most central collisions and at about 1.1~fm for the most peripheral ones.
A~temperature of about 250~MeV is the smallest temperature, still larger than $T_c$, 
where we expect the hierarchy \eqref{hierarchy} to be safely realized for the bottomonium lower states.
The outcome of the evolution equations depends on the initial condition. 
Both the singlet and the octet are initialized in a Dirac-delta state as they are assumed to be produced by a hard (local) process.
We call $\delta/\als(m_b)$ the initial fraction of octets with respect to singlets 
(octet production is $1/\als$ enhanced with respect to singlet production~\cite{Cho:1995ce}).
In the bottomonium case, the time evolutions of $R_{AA}$ for 30-50\% centrality and 50-100\% centrality 
are shown in Figs.~\ref{fig:ev1} and~\ref{fig:ev2} respectively. 
Results are corrected for feed-down effects using the method of~\cite{Strickland:2011aa}. 
Note that in Fig.~\ref{fig:ev1} the $R_{AA}$ for the $2S$ state becomes insensitive to $\kappa$ at large times, 
an indication that it reaches a steady state before the quark-gluon plasma vanishes.

\begin{table}[h]
\begin{center}
\begin{tabular}{|c|c|c|c|}\hline 
\multicolumn{2}{|c|}{30-50\% centrality}&\multicolumn{2}{|c|}{50-100\% centrality} \\
\hline
$R_{AA}(1S)$ & $\frac{R_{AA}(2S)}{R_{AA}(1S)}$ & $R_{AA}(1S)$ & $\frac{R_{AA}(2S)}{R_{AA} (1S)}$ \\
\hline
$0.23^{+0.10}_{-0.07}$ & $0.24\pm 0.09$ & $0.80 \pm 0.05$ & $0.59 \pm 0.10$ \\
\hline
\end{tabular}
\end{center}
\caption{Results for $R_{AA}(1S)$ and $R_{AA}(2S)$ for $\kappa/T^3$ in the range \eqref{kapparange}, $\gamma = 0$ and $\delta =1$ in the bottomonium case.
\label{tab:results}}
\end{table}

We have taken $\kappa/T^3$ in the range \eqref{kapparange}, while we have set $\gamma = 0$ and $\delta =1$.
The choice of $\gamma$ is arbitrary, since a non-perturbative determination of this parameter is missing.
The CMS results of~\cite{Chatrchyan:2012lxa} prefer lower values of $\gamma$, which is the rational for our choice of $\gamma = 0$. 
The choice $\delta =1$ assumes the initial ratio of octets over singlets to be just $1/\als(m_b)$.
The precise value of $\delta$ is not so important as the result is rather insensitive to it. 

\begin{figure}[h]
\includegraphics[scale=0.35]{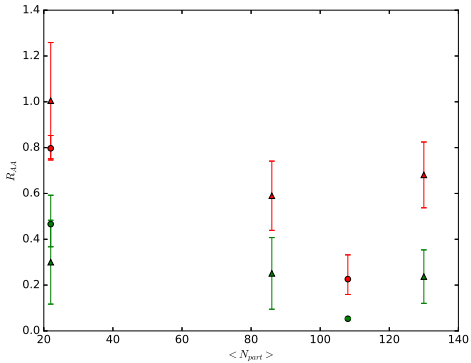}
\caption{$R_{AA}$ as obtained from Tab.~\ref{tab:results} (dots) compared with the CMS data of~\cite{Chatrchyan:2012lxa} (triangles).
Upper (red) entries refer to the $\Upsilon(1S)$, lower (green) entries to the $\Upsilon(2S)$.
\label{fig:RAA}
}
\end{figure}

Our results are summarized in Tab.~\ref{tab:results} and compared in Fig.~\ref{fig:RAA} with the CMS results of~\cite{Chatrchyan:2012lxa}. 
A more recent set of data is in~\cite{Khachatryan:2016xxp}.

\section{Conclusions}
In this letter, we have considered heavy quarkonium in heavy-ion collision experiments as an open quantum system, and derived the evolution equations 
for the singlet and octet density matrices in \eqref{eq:ev}-\eqref{eq:ev_octet} under the assumption that pNRQCD \eqref{eq:Lagrangian_pNRQCD} is applicable, 
which is our main result. For the case of a strongly-coupled plasma under the conditions \eqref{hierarchy}, 
the equations depend only on two non-perturbative parameters, the heavy-quark momentum diffusion coefficient $\kappa$ 
and the parameter~$\gamma$ respectively defined in \eqref{kappa} and \eqref{gamma}, and, furthermore, can be written in the Lindblad form \eqref{eq:Lindblad}. 

We have solved numerically the equations for the case of bottomonium, 
taking the initial conditions according to NRQCD production in vacuum and the evolution of the plasma according to Bjorken's model. 
We obtain the nuclear modification factors $R_{AA}(1S)$ and $R_{AA}(2S)$ (see Tab.~\ref{tab:results}).
The output depends crucially on the parameters $\kappa$ and~$\gamma$.
While $\kappa$ has at least been computed in quenched lattice QCD for temperatures close to $T_c$, $\gamma$ has not, and, therefore,
it remains as one of the main uncertainties in the determination of~$R_{AA}$.

A detailed derivation of the results presented here as well as an analysis of the evolution equations for the case of a weakly-coupled 
quark-gluon plasma will be presented in a forthcoming publication~\cite{Brambilla:2016fire}.

Recently, several works have addressed the computation of quarkonium suppression taking into account the quantum evolution,
the imaginary part of the potential and the conservation of the number of heavy particles~\cite{Akamatsu:2011se,Katz:2015qja},
as it is also the case in this paper. 
However an important difference is that in this work we have taken into account 
both color structures of the quark-antiquark pair, the singlet and the octet. 
This distinguishes it from the abelian case and, as we have shown, has important phenomenological consequences. 
Finally, we emphasize that we have kept the full quantum-mechanical nature of the quarkonium during the whole evolution. 
To be definite we have presented results for Bjorken's model of hydrodynamic evolution and for a particular initial condition. 
Clearly this is not a limitation of the formalism, which allows any hydrodynamic evolution of the medium to be incorporated  
and the initial conditions to be tuned to account for pre-equilibrium states like Glasma (for a recent review see, e.g.,~\cite{Gelis:2012ri}).

\section*{Acknowledgements}
N.B. and A.V. thank Torsten Dahms for many communications.
The work of N.B. and A.V. was supported by the Bundesministerium f\"ur Bildung und Forschung (BMBF) 
through the ``Verbundprojekt 05P2015 - ALICE at High Rate (BMBF-FSP 202) GEM-TPC Upgrade and Field theory based investigations of ALICE physics'' under grant No. 05P15WOCA1
and by the DFG cluster of excellence ``Origin and structure of the universe'' (www.universe-cluster.de). 
The work of M.A.E. was supported by the European Research Council under the Advanced Investigator Grant ERC-AD-267258 and by the Academy of Finland, project 303756.
J.S. thanks N.B. and A.V. for hospitality at TUM. He has been supported by the CPAN  CSD2007-00042 Consolider--Ingenio 2010 program, 
and the FPA2010-16963, FPA2013-43425-P and FPA2013-4657 projects (Spain), and the 2014-SGR-104 grant (Catalonia).

\end{document}